\documentclass[letter, printer]{aa}
\usepackage{lineno}
\usepackage{lscape}
\usepackage{lipsum}
\usepackage[varg]{txfonts}
\usepackage{graphicx}
\usepackage[colorlinks=true, citecolor=blue, urlcolor=blue]{hyperref}
\usepackage{placeins}
\usepackage{color, soul}

\begin{document} 

   \title{Cosmogenic gamma-ray and neutrino fluxes from blazars associated with IceCube events}


   \author{Saikat Das
          \inst{1}
          \and
          Soebur Razzaque\inst{2}
          \and
          Nayantara Gupta\inst{1}
          }

   \institute{Astronomy \& Astrophysics Group, Raman Research Institute, Bengaluru 560080, Karnataka, India\\
              \email{saikatdas@rri.res.in, nayan@rri.res.in}
         \and
             Centre for Astro-Particle Physics (CAPP) and Department of Physics, University of Johannesburg, PO Box 524, Auckland Park 2006, South Africa\\
             \email{srazzaque@uj.ac.za}
             }

   \date{Received YYYY; accepted ZZZZ}

 
  \abstract
   {Blazars constitute the vast majority of extragalactic $\gamma$-ray sources. They can also contribute a sizable fraction of the diffuse astrophysical neutrinos detected by IceCube. In the past few years, the real-time alert system of IceCube has led to multiwavelength follow-up of very high-energy neutrino events of plausible astrophysical origin. Spatial and temporal coincidences of a number of these neutrino events with $\gamma$-ray blazars provide a unique opportunity to decipher cosmic-ray (CR) interactions in the relativistic jets.} 
   {The aim of this work is to test if the $\gamma$-ray blazars associated with the IceCube neutrino events are also sources of ultrahigh-energy (UHE, $E>10^{18}$ eV) cosmic rays.}
   {Assuming that blazars accelerate UHECRs, we calculate the ``guaranteed'' contribution to the line-of-sight cosmogenic $\gamma$-ray and neutrino fluxes from four blazars associated with IceCube neutrino events. We compare these fluxes by the sensitivities of the upcoming $\gamma$-ray imaging telescopes like CTA and by the planned neutrino detectors like IceCube-Gen2.}
   {We find that detection of the cosmogenic neutrino fluxes from the blazars TXS~0506+056, PKS~1502+106 and GB6~J1040+0617 would require UHECR luminosity $\gtrsim 10$ times the inferred neutrino luminosity from the associated IceCube events, with the maximum UHECR proton energy, $E_{p,\rm max}\approx 10^{20}$~eV. Cosmogenic $\gamma$-ray emission from blazars TXS~0506+056, 3HSP~J095507.9 +355101 and GB6~J1040+0617 can be detected by CTA if the UHECR luminosity is $\gtrsim 10$ times the neutrino luminosity inferred from the associated IceCube events, and for $E_{p,\rm max} \gtrsim 10^{19}$~eV.}
   {Detection of cosmogenic neutrino and/or $\gamma$-ray flux(es) from blazars associated with IceCube neutrinos may lead to the first direct signature(s) of UHECR sources. Given their relatively low redshifts and hence total energetics, TXS~0506+056 and 3HSP~J095507.9+355101 should be the prime targets for upcoming large neutrino and $\gamma$-ray telescopes.}

   \keywords{Astroparticle physics --
   	         galaxies: active -- 
             gamma-rays: general -- 
             neutrinos
            }

   \maketitle
%

\section{Introduction}

The origin of high-energy astrophysical neutrinos is still a mystery despite their discovery by the IceCube experiment almost a decade ago~\citep{Aartsen:2013jdh}. The detection of IC-170922A in spatial and temporal coincidence with the blazar TXS 0506+056 has therefore led to speculations that blazars contribute to the diffuse neutrino flux detected by IceCube \citep{IceCube:2018dnn, IceCube:2018cha}. It is the first detected blazar-neutrino association, where the source was found to be flaring in GeV $\gamma$-rays when the high-energy ($E_\nu\sim0.3$ PeV) $\nu_\mu$ created a ``track-like'' event in the South Pole ice. A chance correlation was disfavored at $3\sigma$ C.L. A few weeks after the neutrino alert, the MAGIC telescope detected $\gamma$-rays above 100 GeV from this source for the first time~\citep{MAGIC:2018sak}. Prompted by this observation, a search for time-dependent neutrino signal in the archival data revealed an excess of high-energy neutrino events between September 2014 and March 2015 at $3.5\sigma$ statistical significance. However, the 160-day window was not accompanied by a $\gamma$-ray flare. 

Detection of the IC-170922A event reaffirms the extragalactic origin of the most IceCube astrophysical neutrinos, which is evident from their near-isotropic sky distribution. Since the detection of sub-PeV neutrinos directly indicates cosmic-ray acceleration to at least PeV energies, they are ideal messengers of hadronic processes in astrophysical sources. Blazars, a subclass of radio-loud Active Galactic Nuclei (AGNs), have their relativistic jets pointed towards Earth. They have been long considered as the accelerators of ultrahigh-energy cosmic rays (UHECRs) \citep{Dermer:2009NJPh,  Dermer:2010iz, Murase_2012,  Tavecchio:2013fwa, Oikonomou:2014xea, Resconi:2016ggj, Rodrigues:2017fmu}. Statistical analysis with early IceCube data indicated a weak correlation between the neutrinos, UHECRs, and AGNs~\citep{Moharana:2015JCAP}. 

Other neutrino events, of lower statistical significance, are also identified in spatial coincidence with blazars \citep{Franckowiak:2020qrq, Giommi:2020hbx}, e.g., IC-190730A and IC-200107A, coincident with the direction of blazars PKS 1502+106 and 3HSP J095507.9+355101, respectively. An analysis of archival neutrino events that satisfies the IceCube real-time alert criteria revealed a correlation of IC-141209A with the blazar GB6 J1040+0617. 

The neutrino flux predicted from the IceCube observation of TXS~0506+056 can be produced inside the jet emission region. A synchrotron self-Compton (SSC) model explains well the multiwavelength SED \citep{Keivani_2018, Cerruti_2019, Gao:2018mnu}, with neutrinos produced by a radiatively subdominant hadronic component. While the SED of most blazars can be well explained by leptonic models alone, a hadronic component is required to explain the observed neutrinos. Moreover, if blazars are capable of accelerating UHECRs, a cosmogenic neutrino flux is expected to be observed along their direction for a sufficiently small intergalactic magnetic field. The secondary $\gamma$-rays produced may contribute to the SED at the highest energies, leading to an unattenuated spectrum. Thus there exists a degeneracy in explaining the blazar SED, which can be lifted by future multi-messenger observations. UHE protons, if accelerated, can escape the jet when the escape time scale is shorter than the acceleration time \citep[see, e.g.,][]{2020ApJ...889..149D}. These UHECRs can interact with photons from the CMB and EBL  to produce secondary neutrinos and $\gamma$ rays with a hard spectrum along the line-of-sight~\citep{Essey_2010a, Essey_2010b, Kalashev_2013, Das:2020hev}. In some cases, the line-of-sight $\gamma$-ray flux has shown to improve fits to VHE spectra of BL Lacs \citep{2020ApJ...889..149D}. However, the luminosity requirement in the UHECR spectrum may be challenging for a super-massive black hole~\citep{Razzaque:2012ApJ}.

\begin{figure}
\centering
\includegraphics[width=0.49\textwidth]{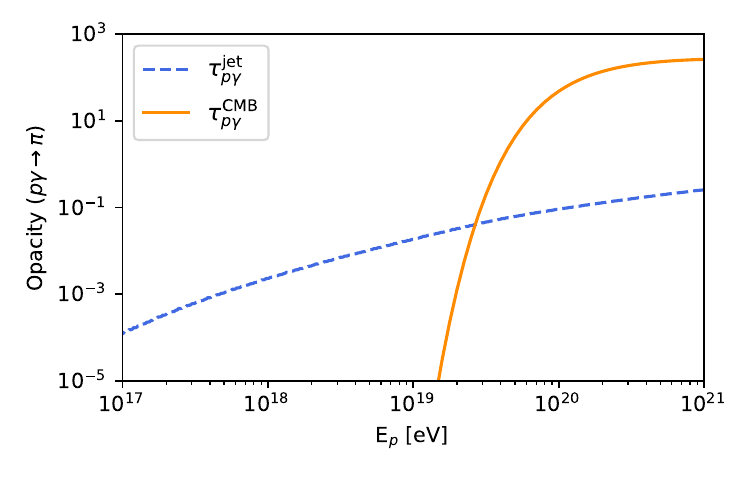}
\caption{\small{Optical depth of photo-pion production inside and outside the source emission region for TXS 0506+056. The target photons for $p\gamma$ interactions inside the jet are leptonic synchrotron and SSC radiation as shown in Fig.~3 of \cite{Gao:2018mnu} for their favored hybrid model. The parameter values given in their Supplementary Table 1, for the flare state, is used to calculate the photon flux in the comoving jet frame.}}
\label{fig:op_dep}
\end{figure}

Cosmic-ray protons with energy $\lesssim 10$ PeV must interact inside the blazar jets in order to produce sub-PeV neutrinos detected by IceCube. The rate for $p\gamma$ interactions in the jet can be smaller than the escape rate, however, and a large fraction of the protons can escape the jet \citep[see, e.g.,][]{2020ApJ...889..149D}. These cosmic-ray protons with energy $\gtrsim 1$ EeV can interact with the CMB and EBL photons to produce photopions. The corresponding $p\gamma$ opacity is $\gg 1$ at energies larger than a few tens of EeV for the distances of the blazars associated with the IceCube events (cf. solid line in Fig.~\ref{fig:op_dep}). In this work, we predict the line-of-sight cosmogenic neutrino flux from a simple scaling relation between the neutrino luminosity inferred from the IceCube event and the injected UHECR luminosity.

Protons with energy $\gtrsim 10$ PeV also interact in the blazar jets, however, the typical $p\gamma$ opacity is $\ll 1$ in the jets (cf. dashed line in Fig.~\ref{fig:op_dep}, see caption for details). Therefore, at energies higher than a few tens of EeV, UHECR protons dominantly interact with the CMB photons and the corresponding neutrino flux, peaking at $\sim 1$ EeV, dominates over any neutrino flux from the jet. Upcoming neutrino detectors such as IceCube-Gen2, GRAND, etc.\ are also most sensitive in the $\sim 1$ EeV range. Hence, we only consider the cosmogenic components for neutrino flux predictions in the UHE range, for the blazars considered in this study. At lower energies the jet component can be important, depending on the flux of CRs that interact inside the jet.

Observations of the EeV neutrinos by the next-generation telescopes with improved sensitivity will set the course for precise modeling of the emission region in the jet. The production of cosmogenic neutrinos will be accompanied by a $\gamma$-ray flux resulting from the electromagnetic (EM) cascade of high-energy secondary $\mathrm{e^\pm}$ and $\gamma$ photons. Future observations by $\gamma$-ray missions, such as CTA and LHAASO, can constrain the flux models and the magnitude of the extragalactic magnetic field (EGMF).

We present the results of our study and discuss them in Sec~\ref{sec:results}. We summarize our results and present an outlook in Sec.~\ref{sec:discussions}. 

\section{Results and Discussions} \label{sec:results}

UHECRs escaping from the blazar jets are deflected by the extragalactic magnetic field (EGMF) from our line of sight. Therefore only a fraction of the secondary neutrinos and $\gamma$ rays, produced by interactions of UHECRs with the CMB and EBL, will be propagating within a narrow cone around the line-of-sight. We calculate this fraction from the distribution of angular deflection of UHECRs on arrival from the blazars associated with IceCube events to the Earth. We assume the UHECR injection spectrum of the form $dN/dE\propto E^{-2.2}$ in the energy range 10 PeV to $E_{p, \rm max}$ and simulate their propagation using CRPropa~3 \citep{Batista_16}, including all relevant energy loss processes. The observer, in this case, is considered to be a sphere of $R_{\rm obs}=l_c$ to increase the number of observed events in the simulation. We incorporate a random turbulent EGMF with Kolmogorov power spectrum of magnitude $B_{\rm rms}=10^{-15}$-$10^{-17}$~G and turbulence correlation length $l_c=0.1$ Mpc, for the extragalactic propagation. The choice of the $B_{\rm rms}$ range is consistent with the lower bound obtained by~\citet{Neronov:2010} from observations of $\gamma$-ray blazars by Fermi-LAT.  

Fig.~\ref{fig:def} shows the distribution of UHECR deflection in the EGMF, binned over $1^\circ$ intervals, for TXS~0506+056 at a redshift of $z=0.3365$ and various $B_{\rm rms}$ values. A conservative limit to the flux of $\gamma$ rays and neutrinos arriving at Earth can be obtained by multiplying the total flux with the UHECR survival fraction within $1^\circ$ of the jet emission direction. For track-like events in IceCube, the angular resolution is 0.5$^\circ$ for a neutrino of energy $E_\nu\sim30$ TeV. We denote this survival fraction within $0-1^\circ$ by $\xi_B$, i.e., the fraction observable at Earth. The value of $\xi_B$ is fairly constant for $E_{p, \rm max}=1$, 10, and 100 EeV, while keeping all other parameters fixed.

\subsection{TXS~0506+056}

\begin{figure}
\centering
\includegraphics[width=0.49\textwidth]{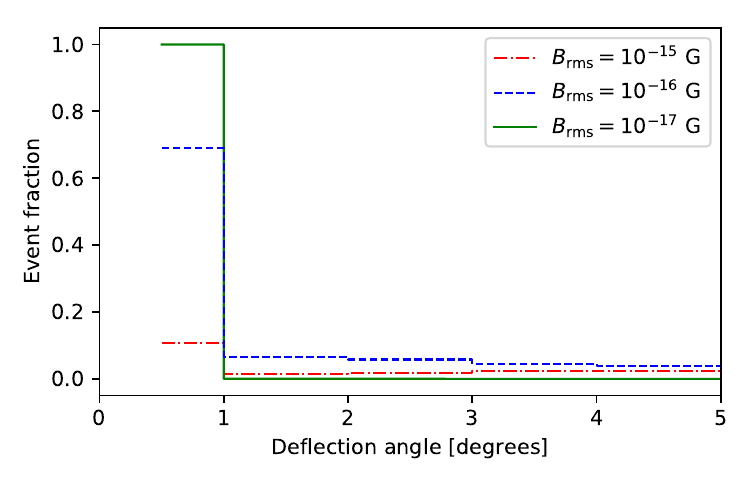}
\caption{\small{Distribution of angular deflection of cosmic rays, over a sphere of radius 100 kpc centered at the observer, resulting from the propagation through extragalactic space for $z=0.3365$}}
\label{fig:def}
\end{figure}
\begin{figure*}
\centering
\includegraphics[width=0.49\textwidth]{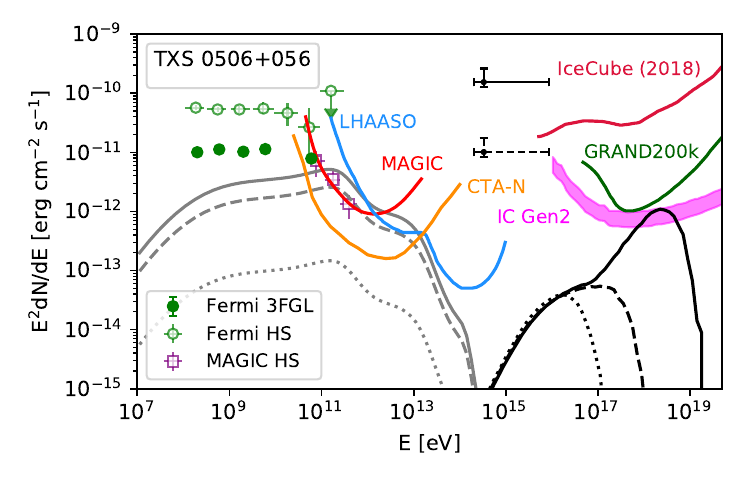}
\includegraphics[width=0.49\textwidth]{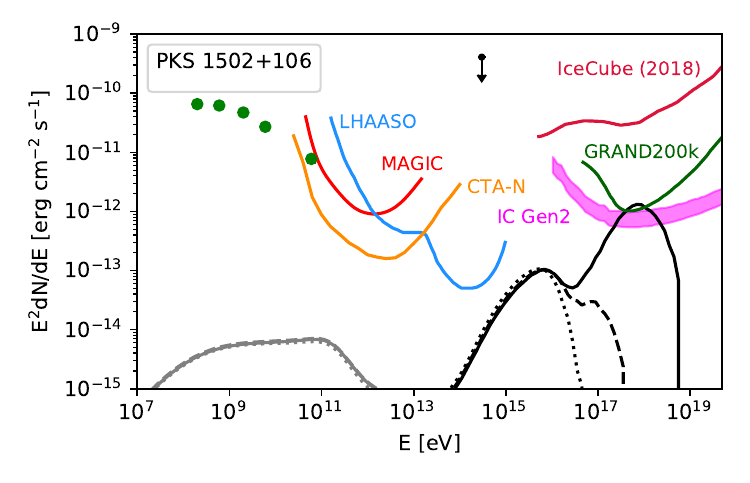}
\includegraphics[width=0.49\textwidth]{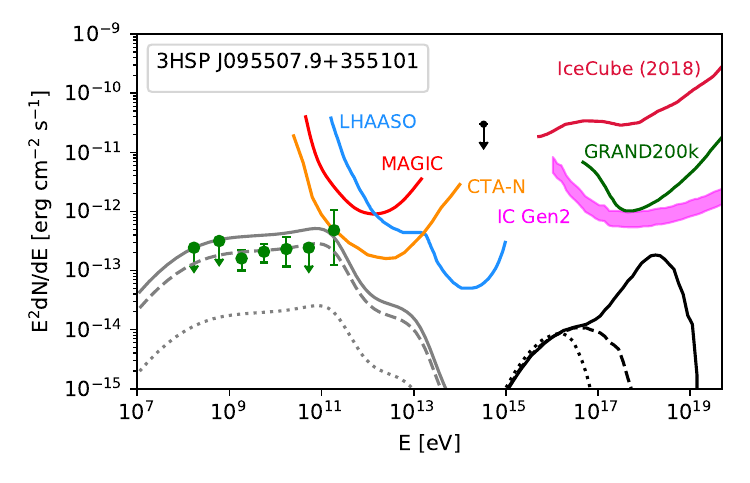}
\includegraphics[width=0.49\textwidth]{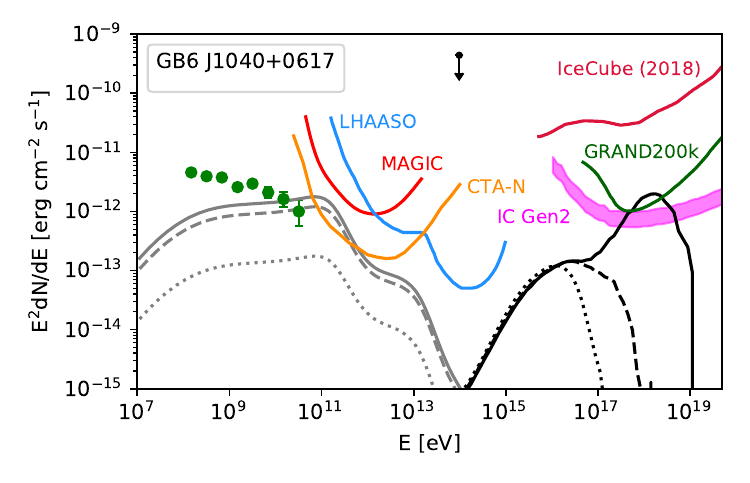}
\caption{\small{Line-of-sight cosmogenic neutrino and $\gamma$-ray fluxes from the TXS~0506+056 (top left) and other blazars (see legend) associated with the IceCube neutrino events. To calculate the cosmogenic fluxes, we assume that UHECRs with energy 10 PeV to $E_{p, \rm max}$ escape from the blazars, where the lower range is consistent with the proton energy required to produce neutrinos observed by IceCube via $p\gamma$ interactions. The solid, dashed, and dotted lines corresponds to $E_{p, \rm max}=1$, 10, and 100 EeV. We assume the UHECR luminosity of the associated blazar is $L_{\rm UHECR} = \alpha L_{\nu}$, where $L_\nu$ is the all-flavor neutrino  luminosity inferred from the track-like events observed by IceCube. The green filled and open circles correspond to $\gamma$-ray flux data points from Fermi-LAT while open magneta squares are MAGIC data points. Also shown in the plots are sensitivities of the current and upcoming $\gamma$-ray and neutrino telescopes, as well as the limit on cosmogenic neutrinos by IceCube.}}
\label{fig:fluxes}
\end{figure*}

The all-flavor neutrino luminosity $L_{\nu}$ is 3 times the  $\nu_\mu+\overline{\nu}_\mu$ flux reported in case of TXS~0506+056 for $\Delta T=0.5$ yrs and 7.5 yrs \citep{IceCube:2018dnn}. The factor of 3 corresponds to an initial production of neutrino flavors $\nu_\mathrm{e}$, $\nu_\mu$, $\nu_\tau$ in the ratio $1:2:0$ from pion decay. In Fig.~\ref{fig:fluxes} top-left panel we show the all-flavor neutrino fluxes by the black solid and dashed data points. We calculate the luminosity of UHECRs by assuming $L_{\rm UHECR} = \alpha L_{\nu}$, where $\alpha$ is a scaling factor that takes into account the escaping UHECR luminosity due to various physical effects. At $E\lesssim10$ PeV, $p\gamma\to n\pi^+ \to e^+ + \nu_e + \nu_\mu + \overline{\nu}_\mu$ interactions inside the jet produces IceCube neutrinos and the neutron escapes as a CR with $\approx 80\%$ of the proton's energy. The three $\nu$-s carry 15\% of the parent proton energy, and hence $\alpha\approx16/3$. In the UHE range, a higher value is possible as protons can escape dominantly, but we consider the conservative lower bound for this study.

We use CRPropa 3 and the EBL model of~\citet{Gilmore_2012} to calculate cosmogenic photon and neutrino fluxes. The secondary electron and photons are propagated via the DINT code that solves the transport equation to yield the respective spectra after taking into account the EM cascade \citep{Lee_98, Heiter_18}. For $\Delta T=0.5$ yrs, $L_\nu=6.2\times10^{46}$ erg/s and for $\Delta T=7.5$ yrs, $L_\nu=4.3\times10^{45}$ erg/s. We show the cosmogenic fluxes for $\Delta T=0.5$ yrs in Fig.~\ref{fig:fluxes}, with $\alpha=16/3$ and  $\xi_B=0.7$ for $B_{\rm rms}= 10^{-16}$~G (see Fig.~\ref{fig:def} ). The data points in green filled circles show the $\gamma$-ray flux in the quiescent state as reported in the Fermi-LAT 3FGL catalog \citep{Fermi-LAT:3FGL}. 

In Fig.~\ref{fig:fluxes} we also show the sensitivities of currently operating and upcoming neutrino telescopes, as well as the flux upper limit from 9 years of IceCube observations \citep{Aartsen:2018vtx}. The projected sensitivity from 5 years of observation by IceCube Gen2 radio upgrade is the shaded region corresponding to the uncertainties in the extrapolation \citep{IceCube:2019pna}. The projected sensitivity of the full 200k radio antenna configuration of the GRAND detector is for 3 years of observation \citep{Alvarez-Muniz:2018bhp}. Assuming the telescopes have the same sensitivity within a zenith angle of $\theta_z$, the solid angle they cover is $\Omega_z = 2\pi(1-\cos\theta_z) \sim 0.38-3.14$ sr for $\theta_z = 20-60$ degree, which is of the order 1. Hence, we plot the actual neutrino flux sensitivity curves reported by the current/upcoming telescope collaborations in units of erg cm$^{-2}$ s$^{-1}$ sr$^{-1}$.

The Cherenkov Telescope Array (CTA) is a ground-based imaging telescope to observe $\gamma$-rays at very-high energies from 20 GeV to hundreds of TeV. A northern and a southern array are currently under construction. In Fig.~\ref{fig:fluxes} we show the predicted differential point-source sensitivity as a function of reconstructed energy in the northern array site of CTA, assuming a 50 hour observation time and pointing to 20 degrees zenith \citep{Gueta:2021vO}. The Large High Altitude Air Shower Observatory (LHAASO) is another air shower observatory that aims to detect $\gamma$-rays up to PeV energies. We show the LHAASO 1-yr sensitivity to Crab-like  $\gamma$-ray point source \citep{Vernetto:2016gro}. The sensitivity of the Major Atmospheric Imaging Cherenkov (MAGIC) telescope corresponding to 50 hrs of observation is displayed \citep{Aleksic:2014lkm}. 

It can be seen from Fig.~\ref{fig:fluxes} that for $E_{p, \rm max} > 10$ EeV the cosmogenic $\gamma$ ray flux exceeds the MAGIC observation \citep{MAGIC:2018sak} of TXS~0506+056 post-IceCube detection in the enhanced VHE $\gamma$-ray emission state (data points shown by violet open squares in the top left panel), if the IceCube neutrino flux is at $\Delta T= 0.5$~yr level.  The MAGIC data points, therefore, serve as upper limits to the line-of-sight $\gamma$-ray flux contribution, in the sub-TeV energy band, if emission from the jet also contributes to the observed flux, which does not show any significant variability in MAGIC data. If CTA detects $\gamma$-ray variability in the multi-TeV energy band, then the cosmogenic origin will be disfavored. The spectral index of the VHE emission in the TeV range will be crucial to understanding the origin. If it matches with the synchrotron, one can expect the yield to be leptonic, whereas gamma rays of hadronic origin are expected to have more cascade interactions inside the jet due to their higher energy. In any case, the jet component is expected to give a softer $\gamma$-ray spectrum after EBL correction, compared to the cosmogenic component. For the cosmogenic component, a hard spectrum is expected at a few TeV energies, as shown in Fig.~\ref{fig:fluxes}. IceCube Gen-2 will be able to detect cosmogenic neutrino flux for $\alpha\gtrsim10$, $E_{p, \rm max}=10^{20}$ eV, $\Delta T= 0.5$~yr flux level. Whereas $\alpha \gtrsim 30$ will be required for detection at $\Delta T= 7.5$~yr flux level.

\subsection{PKS~1502+106, 3HSP~J095507.9+355101 and GB6~J1040+0617}

We do a similar study for a few other IceCube neutrino events, which are significantly associated with blazars. The number of muon neutrino and antineutrino events observed at IceCube in a given operational time $\Delta T$ is given by
\begin{equation}
N_{\nu_\mu} = \Delta T \int_{\epsilon_{\nu, \rm min}}^{\epsilon_{\nu, \rm max}} d\epsilon_\nu \frac{d\Phi_{\nu_\mu}}{d\epsilon_\nu} \langle A_{\rm eff}(\epsilon_\nu)\rangle_\theta 
\label{eqn:nu_flux}
\end{equation}
where $\langle A_{\rm eff}(\epsilon_\nu) \rangle_\theta$ is the effective detector area averaged over the zenith angle bin concerned \citep{Aartsen:2018ywr}. For neutrino flux calculations, we use an updated effective area for event selection as a function of neutrino energy from \cite{Stettner:2019tok}. We assume a muon neutrino flux of the form $d\Phi_{\nu_\mu}/d\epsilon_\nu = k\epsilon_\nu^{-2}$ and an operation time $\Delta T=10$ years, i.e., from when IceCube became fully operational in 2010 up until 2020. The all-flavor neutrino flux can be approximated as three times the $\nu_\mu+\overline{\nu}_\mu$ flux.

The high-energy astrophysical muon neutrino candidate IC-190730A was detected by IceCube with energy $\epsilon_\nu\approx300$ TeV. The Fermi catalog source 4FGL J1504.4+1029, associated with the FSRQ PKS~1502+106 at $z=1.84$, was found to be within 50\% uncertainty region, at an offset of 0.31 degrees from the best-fit neutrino direction \citep{2019GCN.25225....1I}. Assuming this to be the only event from this source in 10 yrs, the estimated all-flavor neutrino flux using equation~(\ref{eqn:nu_flux}) is $\approx 4.1\times10^{-10}$ erg cm$^{-2}$ s$^{-1}$ and hence $L_\nu\approx 10^{49}$~erg~s$^{-1}$. We find the survival fraction of UHECRs, $\xi_B=0.25$ within $1^\circ$ of the initial emission direction towards the earth. 

IC-200107A has been associated with a $\gamma$-ray blazar 3HSP~J095507.9+355101, a BL Lac at $z=0.557$, listed in the Fermi-LAT 4FGL catalog \citep{2020GCN.26655....1I}. It is an extreme blazar with the peak synchrotron frequency $\nu_s>10^{17}$~Hz, located at an angular separation of 0.63 degrees from the best-fit neutrino direction. There is another source, 4FGL~J0957.8+3423, within the 90\% localization region but 1.5 degrees from the neutrino event. The integrated all-flavor neutrino flux needed to produce the single event of $\epsilon_\nu=0.33$ PeV in 10 years is found to be $3\times10^{-11}$ erg cm$^{-2}$ s$^{-1}$ \citep{Giommi:2020viy}. The corresponding all-flavor neutrino luminosity is $L_\nu = 4.1\times10^{46}$ erg s$^{-1}$. For a source at this redshift, we find  $\xi_B=0.52$. 

An archival neutrino event IC-141209A, satisfying the IceCube real-time trigger criteria, was found to be spatially coincident with the $\gamma$-ray source GB6~J1040+0617 \citep{Aartsen:2019gxs}. The chance probability of this coincidence was found to be 30\% after trial correction. The source is listed in the Fermi-LAT 3FGL catalog at $z=0.7351$ and lies within the 90\% uncertainty region of the best-fit neutrino direction. The event deposited 97.4 TeV energy in the IceCube detector. We calculate the survival fraction to be $\xi_B=0.4$ and the all-flavor neutrino flux to be $\Phi_{\nu+\overline{\nu}}=4.4\times10^{-10}$ erg cm$^{-2}$ s$^{-1}$, and thus $L_\nu = 1.1\times10^{48}$~erg~s$^{-1}$.

In Fig.~\ref{fig:fluxes} we show the line-of-sight cosmogenic fluxes for the three blazars discussed above, with $\alpha=16/3$ and the respective UHECR survival fraction $\xi_B$  and three values of $E_{p, \rm max}$, viz., 1, 10, and 100 EeV corresponding to the dotted, dashed and solid lines, respectively. Also shown are the $\gamma$-ray flux data from Fermi-LAT as well as sensitivities of different telescopes. The inferred neutrino luminosities from the associated IceCube events in case of PKS~1502+106 and GB6~J1040+0617 are much higher than the Eddington luminosity of a super-massive black hole, $L_{\rm Edd} = 1.3\times 10^{47}(M_{\rm bh}/10^9 M_\odot)$~erg~s$^{-1}$. Therefore it is unlikely that these are the source of the detected IceCube neutrinos. Because of its high redshift, the cosmogenic $\gamma$-ray flux from PKS~1502+106 is very low. CTA-North will be able to constrain the cosmogenic $\gamma$-ray flux from GB6~J1040+0617, however. The cosmogenic neutrino flux for both the blazars can be tested with IceCube Gen-2 for such extreme luminosity of a blazar. The case of 3HSP~J095507.9+355101 is the most compelling among these three blazars as a plausible IceCube neutrino source. The hard $\gamma$-ray spectrum observed by Fermi-LAT could be explained as cosmogenic $\gamma$ rays for $E_{p,\rm max}\approx 10$~EeV. If a hard spectrum is also detected at a few TeV energies by CTA, it can be the signature of cosmogenic origin, because such hardening may not be seen for intrinsic emission from the jet due to EBL attenuation resulting in a soft spectrum. The sensitivity of CTA-North is comparable to the plotted flux level in Fig.~\ref{fig:fluxes} and will be in a position to constrain the parameter $\alpha$. Detection of cosmogenic neutrino from 3HSP~J095507.9+355101 by IceCube Gen-2 will require $\alpha \gtrsim 10$.

\section{Summary and Outlook} \label{sec:discussions}

In the last few years, there is growing evidence of blazar candidates for several high-energy IceCube events \citep{Franckowiak:2020qrq}. With observations of several well-reconstructed track events by the new Gold-alert system and the electromagnetic follow-up observations, the detection of neutrino events associated with transient phenomena is now possible. The IceCube events considered here have been studied earlier for self-consistent modeling of the blazar SED and emission of neutrinos from the jets~\citep{Banik:2019twt, Petropoulou:2020pqh, Rodrigues:2020fbu}. In some scenarios, a significant fraction of neutrinos at EeV energies can be attributed to that produced from inside the sources \citep{Rodrigues:2020pli}.

We have presented here cosmogenic neutrino and $\gamma$-ray fluxes from these blazars along the line-of-sight, assuming that UHECRs with energy $10^{16}-10^{20}$~eV can be accelerated and can escape from the jets. We use a suitable value of the injection spectral index in the aforementioned energy range, in the absence of an unequivocal choice. The neutrino flux may vary moderately for other values, while the cosmogenic $\gamma$-ray spectrum depends more on the cosmic background photons. For comparison, we also show the neutrino flux for lower values of $E_{p, \rm max}=$ 1 and 10 EeV, while keeping the value of $\alpha=16/3$. In these cases, the cosmogenic neutrino fluxes are practically undetectable.

We used a realistic value of $10^{-16}$~G for the intergalactic magnetic field for deflection of UHECRs from the line-of-sight. The normalization of the cosmogenic fluxes is set by assuming that the UHECR luminosity of the source is a factor $\alpha$ times the sub-PeV neutrino luminosity inferred from the respective IceCube event. The line-of-sight cosmogenic fluxes appear with hard spectra compared to the fluxes directly from the jets and can be identified by the neutrino and $\gamma$-ray telescopes. In case the cosmic rays are not accelerated to $10^{20}$~eV, the cosmogenic neutrino fluxes will be lower than we estimate here. Both higher luminosity in UHECRs and a higher value of $E_{p, \rm max}$ can aid in the detection of cosmogenic neutrinos. But given the sensitivity of future detectors, an observation of cosmogenic neutrino event at $\approx1$ EeV is more likely if the maximum rigidity of CR injection is near to the GZK cutoff energy.

We find that the inferred sub-PeV neutrino luminosities of PKS~1502+106 and GB6~J1040+0617 are already above the Eddington luminosity because of their relatively high redshifts, and a further requirement of a factor $\gtrsim 5$ times this luminosity in UHECRs for cosmogenic neutrino detection by IceCube Gen-2 is unrealistic. Because of the high redshift, cosmogenic $\gamma$-ray flux from PKS~1502+106 is very low. In the case of GB6~J1040+0617, CTA will be able to constrain the cosmogenic $\gamma$-ray flux. 

We also find that the blazars TXS~0506+056 and 3HSP~J095507.9+355101 are very promising sources for the detection of cosmogenic neutrino and $\gamma$-ray fluxes. The required luminosities in UHECRs are $\sim 10$--100 times (for $E_{p,\rm max}\approx 10^{20}$~eV) of the sub-PeV neutrino luminosities inferred from $\sim 10$ years of IceCube data, but are consistent with the Eddington luminosity. Upcoming CTA will be able to detect or constrain the cosmogenic $\gamma$-ray fluxes and shed light on the question of whether sub-PeV neutrino sources are also the sources of UHECRs.   

\begin{acknowledgements}
S.D. thanks Kumiko Kotera and Matteo Cerruti for helpful discussions. The work of S.R. was partially supported by the National Research Foundation (South Africa) and by a University of Johannesburg Research Council grant. 
\end{acknowledgements}

%
%

\bibliographystyle{aa} 
\bibliography{aa.bib}

\end{document}